\documentclass[universe,review,accept,pdftex]{Definitions/mdpi} 

\firstpage{1} 
\pubvolume{7}
\issuenum{5}
\articlenumber{132}
\pubyear{2021}
\copyrightyear{2021}
\history{Received: 26 March 2021; Accepted: 28 April 2021; Published: 4 May 2021}

\pdfoutput=1

\usepackage{subfigure}
\usepackage{amssymb}
\usepackage[italicdiff]{physics}
\usepackage{natbib}[maxcitenames=3]
\usepackage{booktabs,caption}
\usepackage[flushleft]{threeparttable}
\setcitestyle{authoryear,open={(},close={)}}

\Title{Properties of Fossil Groups of Galaxies}

\Author{J. Alfonso L. Aguerri $^{1,2,\dagger}$\orcidA and Stefano Zarattini $^{3,4,\dagger}$\orcidB}

\AuthorNames{J. Afonso L. Aguerri and Stefano Zarattini}

\address{%
$^{1}$ \quad Instituto de Astrof\'isica de Canarias; C/ Vía L\'actea s/n, 38200, La Laguna, Spain; e-mail: jalfonso@iac.es\\
$^{2}$ \quad Departamento de Astrof\'isica, Universidad de La Laguna, E-38206 La Laguna, Spain\\
$^{3}$ \quad Dipartimento di Fisica e Astronomia ``G. Galilei'', Universit\`a di Padova, vicolo dell'Osservatorio 3, I-35122 Padova, Italy; e-mail: stefano.zarattini@unipd.it\\
$^{4}$ \quad INAF–Osservatorio Astronomico di Padova, vicolo dell’Osservatorio 2, I-35122 Padova, Italy}

\firstnote{The authors contributed equally to this work.} 

\abstract{We review the formation and evolution of fossil groups and clusters from both the theoretical and the observational points of view. In the optical band, these systems are dominated by the light of the central galaxy. They were interpreted as old systems that had enough time to merge all the M* galaxies within the central one.
During the last two decades many observational studies were performed to prove the {\it old and relaxed} state of fossil systems. The majority of these studies, that spans a wide range of topics including halos global scaling relations, dynamical substructures, stellar populations, and galaxy luminosity functions, seem to challenge this scenario.
The general picture that can be obtained by reviewing all the observational works is that the fossil state could be transitional. Indeed, the formation of the large magnitude gap observed in fossil systems could be related to internal processes rather than an old formation.}


\begin{document}


\section{Introduction}
The Lambda cold dark matter scenario ($\Lambda$CDM) predicts that structures in the Universe form following a hierarchical evolution: small objects collapsed first under their self-gravity and are then merged continuously to build larger structures. In this scenario, galaxy formed first, then merged in small groups and the process continue until the creation of massive galaxy clusters \citep{White1978,White1991}.

\citet{Ponman1993} firstly suggested, while studying compact groups, that this building scenario could be taken to the extreme consequences. They predicted that, in some cases, all the main galaxies of a group could merge with one another, creating a giant galaxy embedded in an X-ray halo typical of a group. This prediction was supported, one year later, by the discovery of RX 11340.6+4018, an apparently isolated elliptical galaxy, at redshift $z=0.171$, found in an extended X-ray halo. The estimated X-ray mass was $2.8 \times 10^{13}$ M$_\odot$, making it a typical group-sized object \citep{Ponman1994}. These systems were named as ``{\it fossil groups}'' (FGs). They were thought to be the latest stage in the evolution of galaxy groups. For this reason they were supposed to be old and dynamically relaxed systems \citep{Ponman1993}. 

The spatial density and fraction of FGs were estimated in different studies, both theoretically and observationally. The agreement on the spatial density is quite good, with estimations in the range $\sim 1-3 \times 10^{-6} \,h_{50}^{3}$ Mpc$^{-3}$ \citep[see e.g. ][]{Vikhlinin1999,Santos2007,LaBarbera2009}. On the other hand, the fraction of FGs with respect to the total amount of clusters and groups is more debated. A comparison between different works can be found in table 1 of \citet{Dariush2007}: the estimated fractions varies between 1\% and 40\%, with a large scatter, and a mean value can be found at about $\sim 10-15\%$. 

In this review we show the main observational and theoretical works related to FGs written during the last three decades. Our aim is to show that these systems fix well in the current theory of structure formation in the Universe. They are just extreme systems produced following this theory. This review is structured in the following way: in Sect. \ref{sec:search} we will describe the search for FGs in the last $\sim 20$ years and their observational definitions. In Sect. \ref{sec:theory} we will present the theoretical scenarios proposed to describe these systems. In Sect. \ref{sec:observations} we will discuss the properties of the intra-cluster medium and of the galaxy populations. Then, we will describe  the possible progenitors of FGs in Sect. \ref{sec:other}. Finally, in Sect. \ref{sec:conclusions} we will propose a sample of ``{\it genuine}'' FGs and draw our conclusions.

\section{The search for fossil systems}
\label{sec:search}

\begin{figure}
    \centering
    \includegraphics[width=\textwidth]{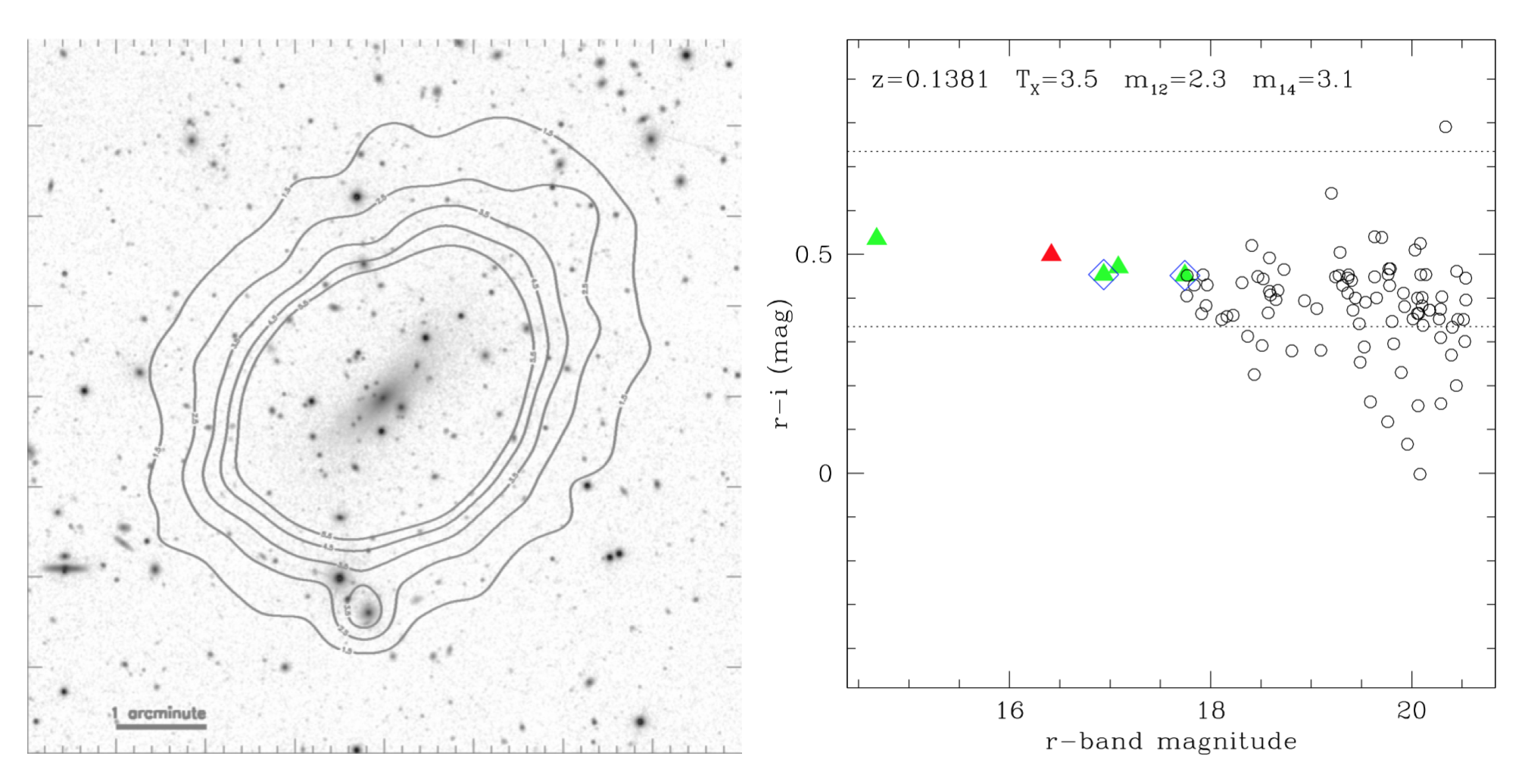}
    \caption{Left panel: SDSS image with X-ray contours of Abell 1068. Right panel: color-magnitude diagram of the same object. Green triangles are system members, whereas red triangles are those that are not. The redshift, $\Delta m{12}$, and $\Delta m{14}$ are reported in the right panel. The original image can be found in \citet{Harrison2012}, in particular it corresponds with their fig. B7.}
    \label{fig:FG_red_sequence}
\end{figure}

Since their discovery in 1994, a strong observational effort was done to find FGs. Despite their supposed frequency, more than 10 years were needed before producing reasonably large catalogues of some tens of candidates. In this sense, one of the main obstacles for the community was to agree on a practical definition. In fact, at the beginning the search was limited to isolated galaxies surrounded by an X-ray halo, but this definition was too loose \citep{Vikhlinin1999}. More rigorous definitions, involving photometry, spectrocopy and X-ray data, were proposed and are actually used. We will discuss them in Sect. \ref{sec:definition}. However, these rigorous definitions had important observational limitations. The result of these constraints was that fossil samples were not selected in a homogeneous way and, in Sect. \ref{sec:catalogues}, we will discuss the strength and weaknesses of the different approaches adopted in the literature. 

\subsection{Operational definitions of fossil groups}
\label{sec:definition}

The most common operational definitions of FGs are based on the magnitude gap between the brightest member galaxy of the group/cluster and other galaxy sorted by their magnitude ($\Delta m_{1,j}$, where $j$ represent the j$-th$ ranked galaxy). In particular, \citet{Jones2003} suggested that a group or a cluster of galaxies should be classified as fossil if the magnitude gap between its two brightest members ($\Delta m_{12}$) is larger than 2 magnitudes in the $r-$band. In addition, they also imposed that the two brightest galaxies should be located within half the (projected) virial radius (defined as $r_{200}$, the radius of a sphere which mean density is 200 times the critical density of the Universe). Another common definition is the one presented in \citet{Dariush2010}: a group/cluster of galaxies is classified as fossil if the magnitude gap between its first and fourth brightest member galaxies ($\Delta m_{14}$) is larger than 2.5 magnitudes in the $r-$band and within half the (projected) virial radius. Both definitions also require the presence of a diffuse X-ray halo, with $L_X \ge 10^{42} \,h_{50}^{-2}$ erg s$^{-1}$. This last criterium ensures that the system is located in a potential well similar in mass to groups or clusters of galaxies. An example of a typical FG and the color-magnitude diagram of its galaxy population is shown in Fig. \ref{fig:FG_red_sequence}.

The original name and definition were though to describe only galaxy groups. However, many studies \citep[e.g. ][]{Sun2004,Khosroshahi2006,Aguerri2011} found the existence of fossil clusters. This is due to the lack of an upper limit for the X-ray luminosity in both the most-common operational definitions. As a consequence, when describing the general properties of the population, {\it fossil groups} and {\it fossil clusters} are interchangeable in the literature. Besides, a more general {\it fossil systems} is also widely used. We invite the reader to consider these three definitions as equivalent along this review.

\subsection{Catalogues}
\label{sec:catalogues}

We already mentioned that the prototype of FGs is RX 11340.6+4018, presented in \citet{Ponman1994}. A few years later, \citet{Vikhlinin1999} studied a sample of four {\it X-Ray Overluminous Elliptical Galaxies}: the authors claimed that these objects could have been part of the fossil category, but it was just a suggestion, since no operational definition was available until \citet{Jones2003} proposed the use of the magnitude gap as the discriminating factor between fossils and non-fossils, proposing a sample of 5 FGs in their work.

Other studies presented small numbers of FGs candidates \citep[][ amongst others]{Yoshioka2004,Khosroshahi2004,Sun2004,Ulmer2005,MendesdeOliveira2006,Khosroshahi2006}, whereas the first large sample of FG candidates was presented in \citet{Santos2007}: in this work, the authors selected 34 galaxy aggregations obtained from the Sloan Digital Sky Survey Data Release 5 \citep[SDSS DR5,][]{Adelman-McCarthy2007} by cross matching the sample of SDSS {\it Luminous Red Galaxies} \citep{Eisenstein2001} with sources in the ROSAT all-sky {\it bright source catalogue} \citep{Voges1999}. The result of the cross match was a list of elliptical galaxies surrounded by an X-ray halo. Then, the magnitude gap was computed within a fixed radius (500 kpc) and in a fixed redshift range ($\Delta z = 0.002$ when spectroscopic redshift is available, $\Delta z = 0.035$ when only photometric redshift is known). These 34 FG candidates were then studied in detail by the Fossil Group Origins (FOGO) project \citep{Aguerri2011}. This observational project produced a set of results analysing the properties of these systems. In the framework of this project, \citet{Zarattini2014} confirmed that $15^{+8}_{-5}$ of the candidates are FGs according to the \citet{Jones2003} or \citet{Dariush2010} definitions. The uncertainties in the number of confirmed FGs reflect those on the definition of $r_{200}$. 
The large difference between the proposed candidates and the confirmed fossils can be explained with a stricter implementation of the definition criteria (e.g. differences in the search radius and membership definition). From the comparison between \citet{Santos2007} and \citet{Zarattini2014} it seems clear that three types of observations are needed in order to strictly define an FG: (i) X-ray data, required to estimate the mass of the system and define the virial radius, (ii) multi-object spectroscopy, in order to identify the real members of the FG, removing fore- and background objects, and (iii) an optical image, needed to measure the magnitude of each galaxy and to compute the magnitude gap between the brightest galaxy and the other members. The combination of the strong observational effort together with the demanding observational definition is probably the main limit for the building of a large and homogeneous dataset of fossil systems.
However, the number of known FGs kept growing in the last decade. Without intending to be exhaustive, \citet{LaBarbera2009} used SDSS-DR4 and ROSAT to build a sample of 25 FG candidates, \citet{Tavasoli2011} found 109 FGs obtained from SDSS-DR7 using a friend-of-friend algorithm, \citet{Makarov2011} presented a catalogue of 395 nearby groups and claimed that $\sim 25\%$ of those are FGs, \citet{Harrison2012} presented a sample of 17 FG candidates, and \citet{Gozaliasl2014} presented a catalogue of 129 groups, of which 22 $\pm 6\%$ are fossils. 
While the number of candidates is rising, it become more and more complicated to have dedicated multi-object spectroscopic observations to constraint membership. The main drawback of this strategy is that the purity of the sample is difficult to control and non-fossil systems can dilute the statistical relevance of the results. 
It is thus complicated to estimate the number of ``{\it genuine}'' FGs, namely those that strictly accomplish the operational definitions, known up to date. However, in the last section of this review we will try to define a sample of such genuine systems.

\begin{figure}
    \centering
    \includegraphics[width=\textwidth]{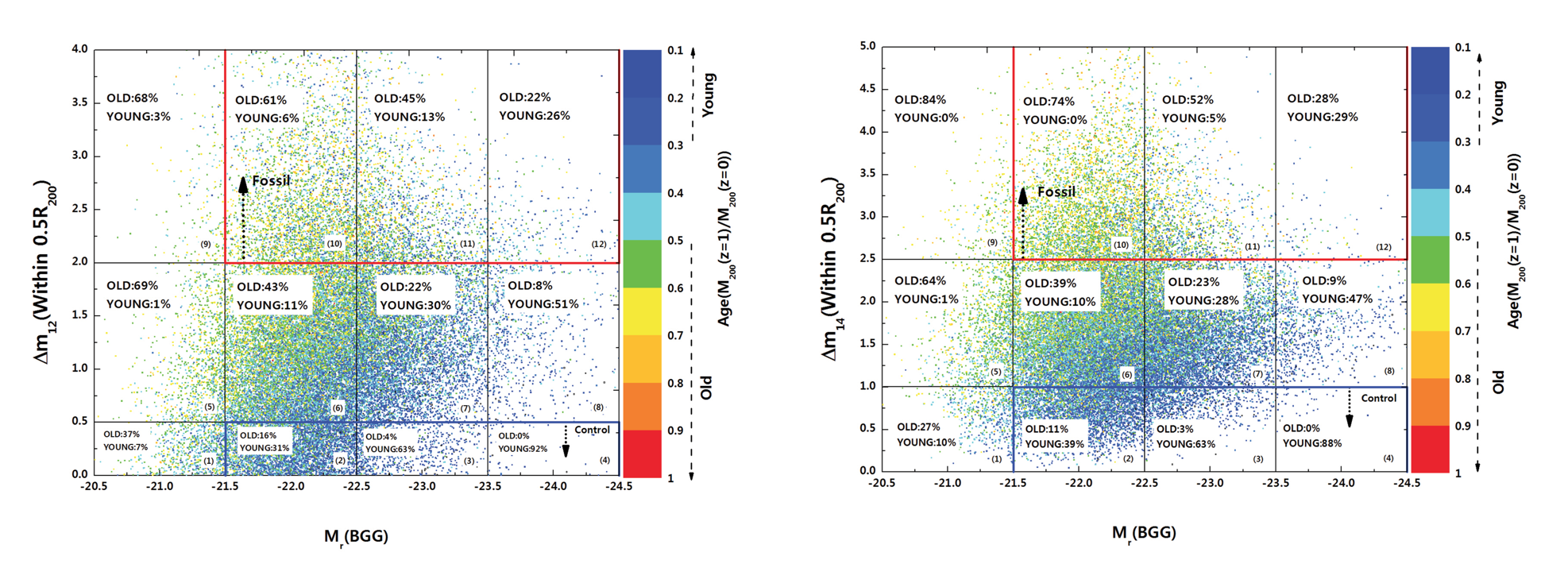}
    \caption{Distribution of the galaxy groups in the plane of luminosity gap $\Delta m_{12}$ (left panel) and $\Delta m_{14}$ (right panel) within 0.5 $r_{200}$ and the $r-$band magnitude of the brightest group galaxy, in the Millennium simulations with Guo et al. (2011) semi-analytic model. Data point are colour-coded according to the ratio of the group halo mass at redshift $z \sim 1$ to its mass at $z = 0$. The plane has been sub-divided into blocks within which the probability that the halo is old or young is given. In this diagram, panels (5), (9) and (10) contain mostly old systems while the panels (3), (4) and(8) are mostly occupied by young systems. The image is taken from \citet{Raouf2014}, in particular it corresponds to their fig. 1.}
    \label{fig:age_fgs}
\end{figure}

\section{Theoretical framework}
\label{sec:theory}
Numerical simulations of the number of satellites in galaxy groups predict the presence of a huge number of dwarf galaxies \citep[e.g. ][]{Moore1999}. However, these predictions are not confirmed in observations. In fact, if the number of bright galaxies is in agreement with these predictions, the number of observed dwarfs is up to two orders of magnitude lower than expected \citep[e.g. ][]{Mateo1998}. This is the so-called {\it missing satellite problem}.
\citet{D'Onghia2004} pointed out that FGs could scale up the missing satellite problem at the mass scales of more massive galaxies. In particular, they compared the number of satellites predicted by $\Lambda$CDM at different halo mass scales and  concluded that FGs shown smaller number of galaxies such as the Milky Way and the Large Magallanic Cloud than those predicted by the structure formation theory. They claimed that the reason of this lack of bright galaxies was due to over merging processes occurred in FGs. These early findings made FGs extreme objects in the structure formation of the Universe. The over merging processes in FGs could be explained in terms of differences in the orbital structure between fossil and non-fossil systems. \citet{Sommer-Larsen2006} used cosmological TreeSPH simulations to study the orbital structure of the intra-group (IG) stars for a set of clusters with masses $\sim 10^{14}$ M$_{\odot}$. He concluded that the velocity distribution of the IG stars was significantly more radially anisotropic for fossil than for non-fossil systems. This pointed out that the initial velocity distribution of the group galaxies could play an important role in defining the fossil status of the system.

Several works have focused on the theoretical study of the mass assembly history of fossil and non-fossil systems. One of the first papers on this topic was done by \citet{D'Onghia2005} where they analysed the mass assembly of systems with $M_{vir} \sim 10^{14}$ M$_{\odot}$. They found a correlation between $\Delta m_{12}$ and the formation time of the group defined as the redshift at which 50$\%$ of the total mass of the system at $z = 0$ is already in place ($z_{50}$). In particular, they found that FGs have assembled more than half of their present mass at $z > 1$, with a subsequent growth by minor mergers alone. This early assembly gave to FGs enough time to merge their M$^{*}$ galaxies (where M$^*$ is the characteristic magnitude of the luminosity function, see Sect. \ref{sec:LFs} for details) and produce the large magnitude gaps observed at $z=0$. The mass assembly of the low-mass group regime ($M_{vir} \sim 10^{13} - 10^{13.5} M_{\odot}$) was analysed by using the Millenium Simulation by \citet{Dariush2007,Dariush2010}. In particular, these authors found that the selection of the systems using their magnitude gap alone does not guarantee the selection of early formed systems. They observed that the majority of the objects that have assembled more than 50$\%$ of their halo mass at $z=1$ are not fossil systems today. A similar result was found in \citet{Deason2013}: 20$\%$ of the groups selected from the Millenium Simulation with large mass gap (similar to fossil systems) turned to be young objects. \citet{Raouf2014} proposed that a combination of three observational parameters (magnitude gap, luminosity of the brightest cluster galaxy and its offset from the group luminosity centroid) considerably improve the selected rate of dynamically old systems. In Fig. \ref{fig:age_fgs} it can be seen that the probability for a system with $\Delta m_{12} > 2.0$ and $\Delta m_{14} > 2.5$ to be old grows when the absolute magnitude of the BCG is smaller. In this case, \citet{Raouf2014} defined a group as old if its halo has over 50\% of its final mass at $z = 1$ and young if this fraction is less than 30\%. Indeed, this plots demonstrate that $\Delta m_{14}$ works better than $\Delta m_{12}$, when combined with the absolute magnitude of the central galaxies, in finding old systems. This result was recently confirmed in \citet{Zhoolideh2020}, since the authors found a clear correlation between $\Delta m_{12}$, the offset of the luminosity centroid, and the dynamical age of a group/cluster. However, this criterium is less used in the literature, probably because it is newer than the \citet{Jones2003} and \citet{Dariush2010} ones.

\begin{figure}
    \centering
    \includegraphics[width=\textwidth]{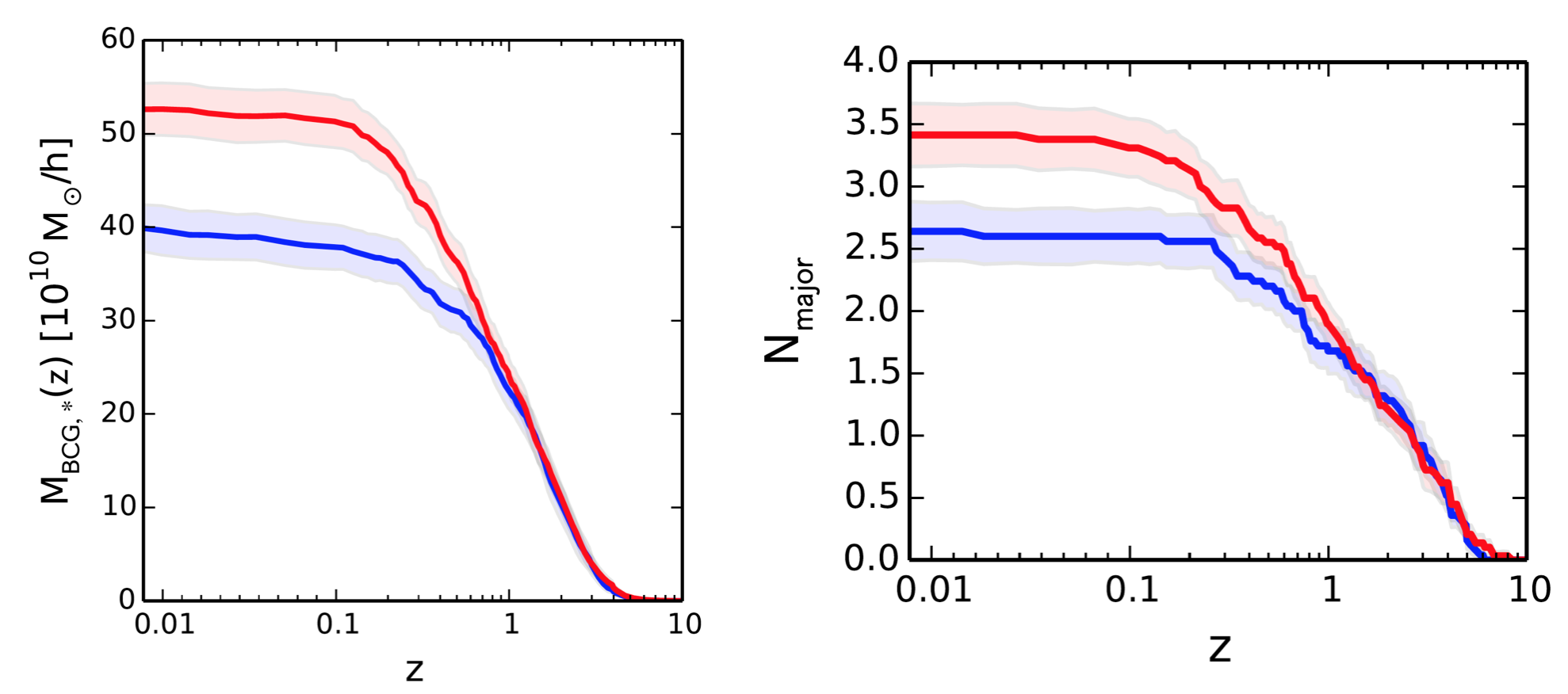}
    \caption{Left panel: average stellar mass assembly history for central galaxies within $r_{200}$. Shaded area are 1$\sigma$ errors calculated from 1000 bootstrap resamplings. Right panel: average number of cumulative major mergers occurred at the central galaxy across z. In both panels, fossil and non-fossil systems are represented in red and blue, respectively. The image is taken from \citet{Kundert2017}, in particular from their fig. 6 and 7.}
    \label{fig:mass_assembly}
\end{figure}

Also the Illustris Simulation was used to analyse the properties of the FGs in the mass regime $10^{13} - 10^{13.5} M_{\odot}$ \citep{Kundert2017}. The authors found that the magnitude gap of FGs identified at $z=0$ were on average created about 3 Gyr ago. In addition, the fossil central galaxies became more massive than non-fossil ones.  This difference was explained as due to differences in the mass acquired through mergers between $z = 0.1 - 1$, as can be seen in the left panel of Fig. \ref{fig:mass_assembly}. Fossil BCGs have also a larger number of major mergers than non-fossil ones (right panel of Fig. \ref{fig:mass_assembly}). Indeed, the last major merger of fossil BCGs was later (e.g. at lower redshift). No differences were found in the distribution of the time formation ($z_{50}$) of fossil and non-fossil halos. The group mass assembly of fossils and non-fossils differs only in the recent group accretion history, in particular in the formation time of the 80$\%$ of the mass of the halo. However, semi-analytical models studied  the dynamical evolution of galaxies in groups with different formation epochs \citep{Raouf2018}. They found that BCGs of dynamically-young groups suffered the last major galaxy merger $\sim 2$ Gyr more recently than their counterparts in dynamically old groups and that FGs are somewhere in the middle between the other two populations.  However, these authors found a lack of recent major mergers in FGs, that is in agreement with the evolution of their old systems.

\citet{Raouf2016} also used the Illustris Simulation to study some properties  of fossil systems. They found that this simulation overproduce FGs in comparison with observations and semi-analytical predictions. They also obtained that the intra-group medium (IGM) in dynamically evolved groups is hotter, for a given halo mass, than that in  still evolving ones.

Several studies found that the fossil phase of a system could be transitional. Galaxy clusters and groups pass through fossil and non-fossil phases along their evolution. \citet{vonBenda2008} found a population of groups that presented a fossil phase at high redshift which is terminated later by the accretion of new bright galaxies. The transitional phase of the fossil status is also reported by other simulations like \citet{Kundert2017}. This fluctuations in the magnitude gap could be related with the large-scale environment in which the systems are located. Indeed, Diaz-Gimenez et al. (2011) found that, in the Millenium Simulation, the environment was different for fossil and non-fossil systems with similar masses. They showed an increase in the local density profile of galaxies at $\sim 2.5 \, r_{vir}$ from the group centers. This increment was more noticeable in fossil than in non-fossil systems and was linked with the earlier formation time of fossil groups. We will discuss in Sect. \ref{sec:other} what is found in observations that can be linked to the transitional fossil phase.

The properties of the galaxy populations in fossil and non-fossil systems have been also analysed using cosmological simulations. \citet{Romeo2015} reported from cosmological hydrodynamical simulations and semi-analytical models that fossil and non-fossil systems show different star forming rates at low $z$, being indistinguishable at $z > 0.5$. In contrast, \citet{Kundert2017} found no differences in the stellar age, metallicity and star formation rates of BCGs in fossil and non-fossil systems from the Illustris Simulation. \citet{Raouf2016} analysed the properties of the black holes developed in the center of the BCGs for fossil and non-fossil systems. They found that the mass of the black holes hosted in BCGs is larger in dynamically evolved groups with a lower rate of mass accretion, a result confirmed also in \citet{Khosroshahi2017}.

\citet{Kanagusuku2016} found that fossil and non-fossil systems selected from the Millennium Simulation shown different galaxy populations. In particular, at early times, FGs comprised two large brightest galaxies surrounded by faint ones. At the faint end of the luminosity function, fossil systems turned to be denser at early times that non-fossil ones. This trend reverses at later time and became similar before $z=0$. This was caused by an increase at a constant rate of the number of faint objects in non-fossil systems. In contrast, the number of faint galaxies reached a plateau at $z \sim 0.6$ in FGs, and then grows faster towards $z=0$. The evolution of the galaxy luminosity function as a function of redshift for fossil and non-fossil systems was studied by \citet{Gozaliasl2014}. They build up luminosity functions of galaxy aggregations based on the Millennium Simulation. They found that the bright end of the galaxy luminosity function strongly evolved for fossil systems from $z = 0.5$ to $z = 0$, with changes in M$^{*} \sim 1.2$ mag. This suggest that the mergers of the M$^{*}$ galaxies in fossil systems have a significant impact in the formation of the bright cluster galaxies. In contrast, the faint-end slope of the luminosity function shows no considerable redshift evolution in fossil systems, unlike in non-fossil ones where it grows by 25 - 42 $\%$ towards low redshifts.

\section{Observational properties}
\label{sec:observations}
In this section we review the observational properties of FGs. In Sect. \ref{sec:gas} we describe the properties of the intra-cluster medium, in particular we present global scaling relations, mass and entropy profiles, cool cores, halo concentrations, and metallicities. In Sect. \ref{sec:galaxies} we analyse the galaxy population and, in particular, FGs' luminosity functions, galaxy substructures, stellar populations, central galaxies, and large scale structures. 

\subsection{Properties of the halos}
\label{sec:gas}
The intra-cluster medium (ICM) is the largest baryonic component in galaxy clusters, responsible for $\sim 10\%$ of the total mass of the cluster \citep[e.g. ][]{Rosati2002}. In fact, galaxy formation is inefficient and only  $\sim 10\%$ of the gas is converted in stars and galaxies \citep{Voit2005}, leaving the vast majority adrift in the intra-cluster space. This gas is trapped in the deep potential well of the cluster and heated to X-ray-emitting temperatures through shocks and adiabatic compressions \citep{Ghirardini2019}.

Gravitational collapse predicts tight scaling relations between ICM and cluster mass, according to the so-called self-similar model \citep{Kaiser1986,Bryan1998}. Moreover, cosmological simulations predict that the scaled thermodynamical profiles of galaxy clusters are nearly universal \citep[e.g. ][]{Frenk1999}. For these reasons, the ICM is a powerful tool to study the formation and evolution of galaxy clusters: deviation from the gravitational collapse predictions can be used to investigate non-gravitational physics, such as cooling and feedback from supernovae and active galactic nuclei (AGN).

\subsubsection{Global scaling relations}
\label{sec:scaling}

\begin{figure}
    \centering
    \includegraphics[scale=0.4]{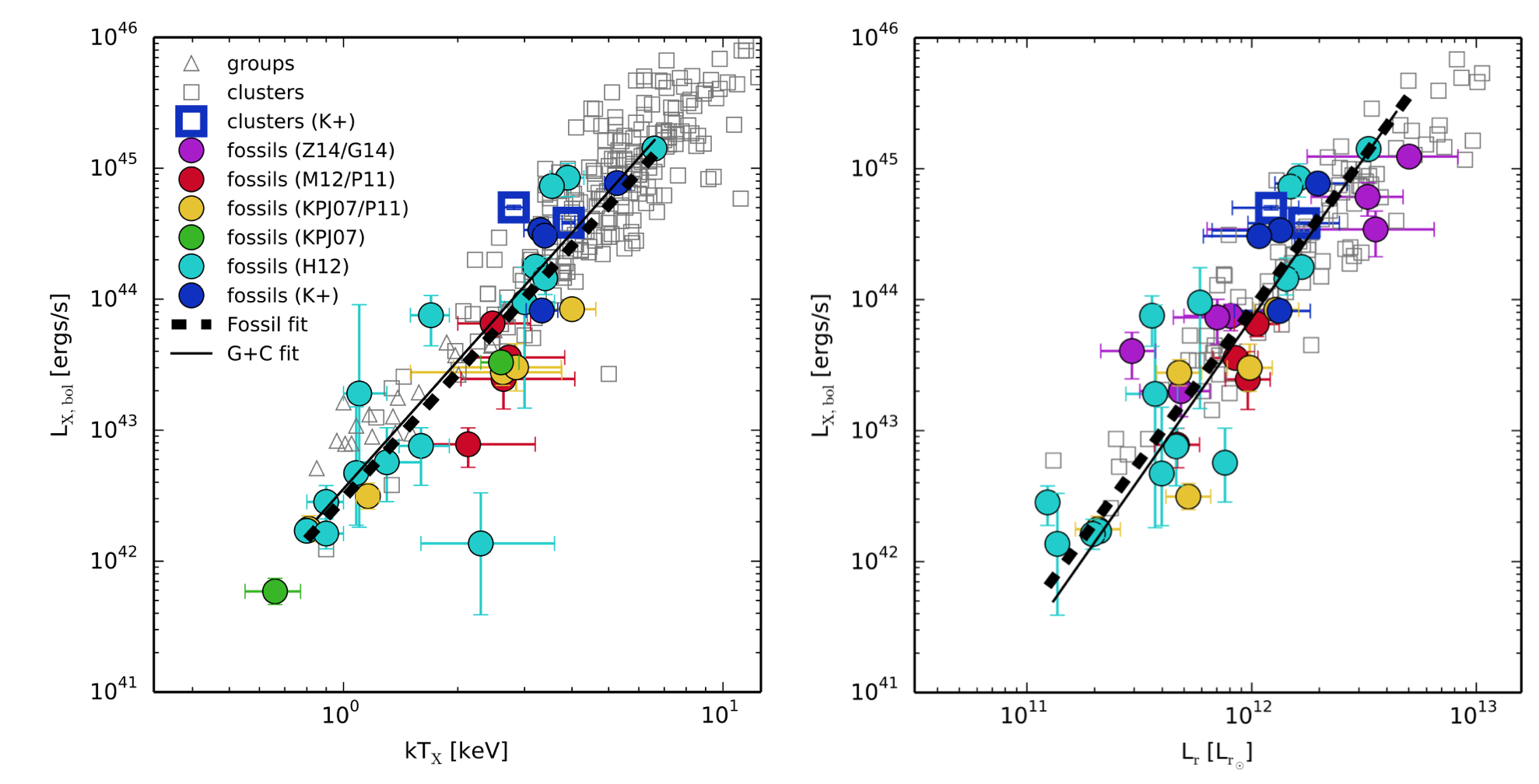}
    \caption{Left panel: The X-ray temperature versus the bolometric X-ray luminosity as presented in \citet{Kundert2015}. Grey triangles and squares are groups and clusters, respectively. Systems labeled with K+ are taken from \citet{Kundert2015}, Z14 from \citet{Zarattini2014}, G14 from \citet{Girardi2014}, M12 from \citet{Miller2012}, P11 from \citet{Proctor2011}, KPJ from \citet{Khosroshahi2007}, and H12 from \citet{Harrison2012}. The plotted lines are the orthogonal BCES fits to the fossil sample (dashed line) and to the sample of groups and clusters (solid line) computed in the same range of parameters. Right panel: the total $r-$band luminosity versus the bolometric X-ray luminosity as presented in \citet{Kundert2015}, with the same lines colour code as in the left panel \citep[see the figure 5 of ][ for details]{Kundert2015}.}
    \label{fig:scaling}
\end{figure}

In the left panel of Fig. \ref{fig:scaling} we show the correlation between X-ray temperatures and luminosities for various samples of FGs and non-FGs. It can be seen that both types of objects are found in the same, tight, correlation. We won't enter in a detailed discussion of purely X-ray relations and their meaning, we refer the reader to the companion review by \citet{Lovisari2021} for a description of X-ray scaling relations in galaxy groups. The result presented in the left panel of Fig. \ref{fig:scaling} can be extended to all those relations that only involve X-ray data: FGs and non-FGs are usually found in the same correlations. This seems to indicate that FGs and non-FGs are formed in a similar way.

However, discussion arose in those studies focused on the comparison between X-ray and optical properties of FGs and non-FGs. In fact, \citet{Jones2003} and \citet{Khosroshahi2007} found FGs to be over luminous in X-rays using samples of five and seven FGs, respectively. They claimed FGs to be a factor 5-10 brighter than regular groups or clusters in the X-ray.

On the other hand, more recent studied found no differences in the statistical properties of FGs and non-FGs samples \citep{Voevodkin2010,Harrison2012,Girardi2014}. In particular, these authors claimed that it is crucial to use homogeneous data and procedures to analyse both the FG and the control samples. According to their statements, this could be the source of the excess of X-ray luminosity (or the lack of optical luminosity) found in the previous studies. 

Nevertheless, \citet{Khosroshahi2014} discussed a sample of groups, one of which defined as fossil, that lies above the $L_{X} - L_{opt}$ relation of non-fossil systems, reopening the debate on fossil system scaling relations.

The (possible) final point of this debate was put by \citet{Kundert2015}: as well as demonstrating that no differences are found between fossils and non-fossils in their sample of 10 groups and clusters observed with specific Suzaku follow up, they recomputed the luminosities of the FGs from \citet{Khosroshahi2007} and \citet{Proctor2011} in a homogeneous way. In the right panel of Fig. \ref{fig:scaling} we show the relation presented in \citet{Kundert2015}. The authors concluded that the discrepancies in the literature can be reconciled if X-ray and optical luminosities were computed using the same bands and radii, pointing out that many differences could be due not only to the low statistics but also the lack of homogeneous datasets.

In parallel with the discussion on the $L_{X} - L_{opt}$ relation, a debate on the mass-to-light (M/L) ratio of FGs arose. In fact, if there is a chance that FGs are under luminous in the optical bands, they should have larger M/L ratios than non-FGs \citep{Khosroshahi2007}. Again, the studies available in the literature are somewhat contradictory. \citet{Sun2004,Khosroshahi2004,Khosroshahi2006} found normal values for their samples of FGs, compatible with non-fossil systems, although if marginally darker for a fixed mass. On the other hand, \citet{Vikhlinin1999} found a high M/L for their sample of  overluminous elliptical galaxies (OLEGs, about 3 times larger). The same result was found by \citet{Proctor2011} in analysing a sample of 10 FG candidates. The authors suggested that FGs are simply dark clusters: they are characterised by a mass and a central galaxy that are typical of galaxy clusters, but embedded in a poor environment, so that the richness and the total optical luminosity are below the non-fossil ones. Finally, \citet{Yoshioka2004} also found high M/L ratios for their sample of ``isolated X-ray overluminous elliptical galaxies''.
However, we suggest that inconsistencies in the measured M/L ratios could be due to differences in the methodology and quality of the data. For example, the optical luminosities of \citet{Vikhlinin1999} and \citet{Khosroshahi2006} are computed in the R band and within $r_{200}$ for every object in the sample, those of \citet{Proctor2011} in a variable radius of $500-1000$ kpc, whereas \citet{Sun2004} estimated a normal M/L ratio from the gas fraction profile out to 450 kpc, and \citet{Yoshioka2004} uses the B-band luminosity of the BCG as the total luminosity of the system.
Thus, it seems reasonable that the disagreement between different results could be healed only with a homogeneous study of a large sample of FGs. As we already mentioned for other topics, such a study is far from being performed.

\subsubsection{Mass and entropy profiles}

The study of mass profiles in clusters is an important tool to confirm the $\Lambda$CDM paradigm. In fact, this model predicts a universal mass profile that does not depend on the mass of the cluster and it is usually assumed to have the shape of a Navarro-Frenk-White profile \citep[NFW,][]{Navarro1997}. 

Mass profiles were studied mainly for individual FGs, or small samples, making difficult to extrapolate general conclusions.
Possibly the first mass profiles of FGs to be computed were those of NGC 6482 \citep{Khosroshahi2004}, RX J1416.4+2315 \citep{Khosroshahi2006}, and ESO 3060170 \citep{Sun2004}. The first two systems shows a mass profile well described by an NFW, with a high central concentration that was interpreted as a sign of early formation. On the other hand, the mass profile of ESO 3060170 showed a flattening in the external regions not compatible with numerical simulations and also confirmed in \citet{Su2013} out to the virial radius. \citet{Yoshioka2004} studied the mass profiles of 4 FG candidates, finding no differences with normal groups/clusters. Also \citet{Gastaldello2007} studied the mass profile of ESO 3060170, within a sample of 16 relaxed groups and clusters: they found a good agreement with a NFW profile, but it must be noticed that their data reached R $ \sim 200$ kpc, whereas \citet{Sun2004} and \citet{Su2013} data reached R $ \sim 500$ kpc and R $ \sim 1000$ kpc, respectively. The three mass profiles are comparable within R $ \sim 200$ kpc and differences with the standard NFW profiles rose in the most external regions. This clarify again the complexity of the comparison when individual FGs are studied using data from different sources, within different radius, and treated with different techniques.
Another two mass profiles for the FGs RXC J0216.7-4749 and RXC J2315.7-0222 were studied in \citet{Democles2010}. Only the latter has a good profile, well fitted, out to $R_{500}$, by a NFW profile plus a central stellar component.

Entropy is also of great interest because it controls ICM global properties and records the thermal history of a cluster, since it is conserved in adiabatic processes. Entropy is therefore a useful quantity for studying the effects of feedback on the cluster environment and investigating any breakdown of cluster self-similarity. Most of the studies cited for the discussion of the mass profiles were also able to compute an entropy profile. Again, since only individual systems were studied, the results are somewhat controversial and it is not trivial to generalise the conclusions to the entire FG category. In particular,  \citet{Democles2010} found that the entropy profiles out to $R_{500}$ of their two FGs show a considerable excess above the expectations from non-radiative simulations, especially for RXC J0216.7-4749. This is expected if significant non-gravitational processes affect the ICM. \citet{Su2013} found and entropy profile that is in agreement with simulations out to $\sim 0.9 \,R_{200}$ and then flattens in the outskirts, due to gas clumpiness and outward redistribution. \citet{Humphrey2012} studied RXJ 1159+5531 combining Suzaku, Chandra, and XMM observations to find no evidence of the flattening in the entropy profile outside $\sim$ R$_{500}$. A similar results was also found in \citet{Su2015} for the same cluster: its entropy profile is consistent with predictions from gravity-only simulations.

The urge for a systematic study of a large sample of FGs in the X-rays appears as necessary to constraint the average mass and entropy profiles of these objects. However, it seems difficult to realise, due to the small number of nearby FGs that can be deeply observed with current X-ray facilities. An improvement on this side is expected with the next all-sky survey that will be taken by eROSITA and Athena missions. The former is expected to find $\sim 10^5$ X-ray clusters and groups, the latter will take advantage of its high-resolution to better constraint radial profiles of, for example, temperature, density, and mass. 

\subsubsection{Cool cores}
Early observations of the gas in galaxy clusters found that it was so dense in the central regions that its cooling time was much shorter than the Hubble time \citep[e.g.][]{Lea1973}. 
The majority of the clusters studied in the literature show these cool cores \citep[CC, e.g.][]{Vikhlinin2005,Hudson2010}. CCs are usually associated to relaxed clusters, since mergers easily erase them. For this reason, it appeared as natural to look for CCs in FGs, in order to confirm their {\it old and dynamically relaxed} status.

The first study on CCs for a sample of FG candidates was done by \citet{Vikhlinin1999}. In their work, the authors studied four isolated elliptical galaxies selected from ROSAT X-ray data and confirmed as fossil systems with a dedicated optical follow up. It is worth noting that the authors suggested that these objects could be FGs, but at the time no operational definition was available, so no $\Delta m_{12}$ is computed in their paper. However, at least three out of four were later confirmed as FGs in other publications, confirming the accuracy of their approach. \citet{Vikhlinin1999} results indicated the presence of CCs in the central regions of these objects.

Later works were focused on deeper studies of individual FGs and found controversial results: \citet{Khosroshahi2004, Khosroshahi2006} found no central drop in NGC 6482 and RX J1416.4+2315. On the other hand, \citet{Sun2004} found a CC in ESO 3060170, as well as \citet{Democles2010} for RXC J0216.7-4749 and RXC J2315.7-0222 and \citet{Su2015} for RXJ1159+5531. 
It is interesting to note that \citet{Miraghaei2014} studied three of the cited clusters (NGC 6482, RX J1416.4+2315, and ESO 3060170) using radio observations, finding signs of recent AGN activity only in the first two. However, the AGN power computed was not sufficient to remove CCs from these clusters.

The need for larger samples was partly satisfied only recently, when \citet{Bharadwaj2016} studied a sample of 17 FGs for which Chandra archival data were available. They defined three different diagnostic to evaluate the presence of CCs and they found that $\sim 80\%$ of FGs showed clear hints of the presence of CCs (e.g. at least two diagnostics compatible with the CC).

It seems, thus, reasonable to claim that FGs are mostly cool-cored. However, the fraction of FGs with a CC is similar to that of non-FGs. For example, \citet{Hudson2010} studied a large sample of 64 galaxy clusters for which high-quality X-ray data from Chandra were available, finding that $\sim 70\%$ of their clusters host a CC. It thus seems that this is not a peculiar behaviour of FGs.

\subsubsection{Halo concentration}
One of the main parameters of the NFW model is the concentration, usually computed as $c_\Delta = r_\Delta / r_s$, where $r_\Delta$ represent the radius of a sphere of mean interior density $\rho_\Delta$ and $r_s$ is the scale radius of the NFW profile. Typical values of $\Delta$ are 200 (often assumed to be equivalent to the virial radius) or 500.

\citet{Navarro1997} pointed out that the concentration parameter reflects the density of the Universe when the halo formed. In particular, older halos formed in higher-density environments and tend to have larger concentrations. Several theoretical studies found that FGs assembled half of their mass at earlier epochs than non-fossil ones \citep[see e.g.][]{Donghia2005, vonBenda2008}. In this framework, it is expected that FGs would be located in high concentrated halos.  Different numerical models are used in the literature to compute halo concentration in clusters \citep[e.g.][]{Dolag2004,DeBoni2013}. Thus, observational results can be easily compared with theoretical predictions to test the formation scenario of FGs.

From an observational point of view, various methods can be used to compute the $c$ parameter. A first approach to derive the central concentration of DM halos is to measure the gas mass profile in the X-rays out to $r_{200}$ or $r_{500}$, fit it with an NFW profile and then use the derived $r_s$ to compute the concentration. Using this approach,  \citet{Khosroshahi2004} \citet{Khosroshahi2006} \citet{Khosroshahi2007}, and \citet{Buote2017} found concentration values higher than expected for their samples of individual FGs 
Other authors, like \citet{Democles2010} and \citet{Pratt2016}, found normal concentrations in a total of 6 FGs. Again, comparing results with such small statistics and taking into account non-homogeneities in the analytic procedures makes difficult to reach a final conclusion that can be applied to the mean FG population. 

However, other approaches can be used to measure halo concentration. In particular, \citet{Vitorelli2018} stacked $\sim 1000$ systems from the CS82 survey \citep{Moraes2014} in different magnitude-gap bins, the larger of which has mean $\Delta m_{12} \sim 1.7$. They cross-correlate weak lensing measurements with NFW parametric mass profiles to measure masses and concentrations of their sample. They found that halos in the $\Delta m_{12} \sim 1.7$ bin have a higher probability to be more concentrated and, thus, probably formed earlier.

Finally, the halo concentration can be estimated using the velocity of member galaxies as tracers of the underlying mass distribution. This was done in \citet{Zarattini2019}, were the authors analysed a sample of $\sim 100$ clusters and groups, dividing them in different magnitude-gap bins. For each bin, they stacked all the available galaxies to increase the statistic. They found $c_{200} = 2.5 \pm 0.4$ for the bin with the largest magnitude gap (defined as $\Delta m_{12} > 1.5$). This values are in agreement, within the uncertainties, with their results in the other three magnitude-gap bins, as well as with other similar work in the literature of non-fossil clusters \citep[e.g. ][ and references therein]{Lin2004,vanderBurg2015}. However, the large uncertainties typical of this observational technique prevented the authors from reaching a strong conclusion.

\subsubsection{Metallicity}
 The hot intra-group gas contains elements that are typically synthesized in stars and SNe. For general details on this topic we refer the reader to the companion review of \citet{Gastaldello2021}. Here, we focus on the single study conducted on metal abundances in FGs, presented in \citet{Sato2010}. The authors get Suzaku data out to 0.5 $r_{180}$ for NCG 1550 and were able to confirm that the abundance ratios O/Fe, Mg/Fe, Si/Fe, and S/Fe are similar to those of other poor groups observed with the same satellite. Moreover, the number ratio of type-I and type-II SNe computed in \citet{Sato2010} is also similar to that obtained for non-fossil groups. As a consequence, their work can be included in those that are not finding differences between FGs and non-FGs.

\subsection{Galaxy population}
\label{sec:galaxies}
The study of the galaxy population in FGs is mainly done in the optical range. In this section, we will firstly discuss the observational properties of the central galaxies. We will then move to the luminosity functions, galaxy substructures, and the large-scale structure around FGs. 

\subsubsection{Central galaxies: formation scenarios}
\begin{figure}
    \centering
    \includegraphics[scale=0.4]{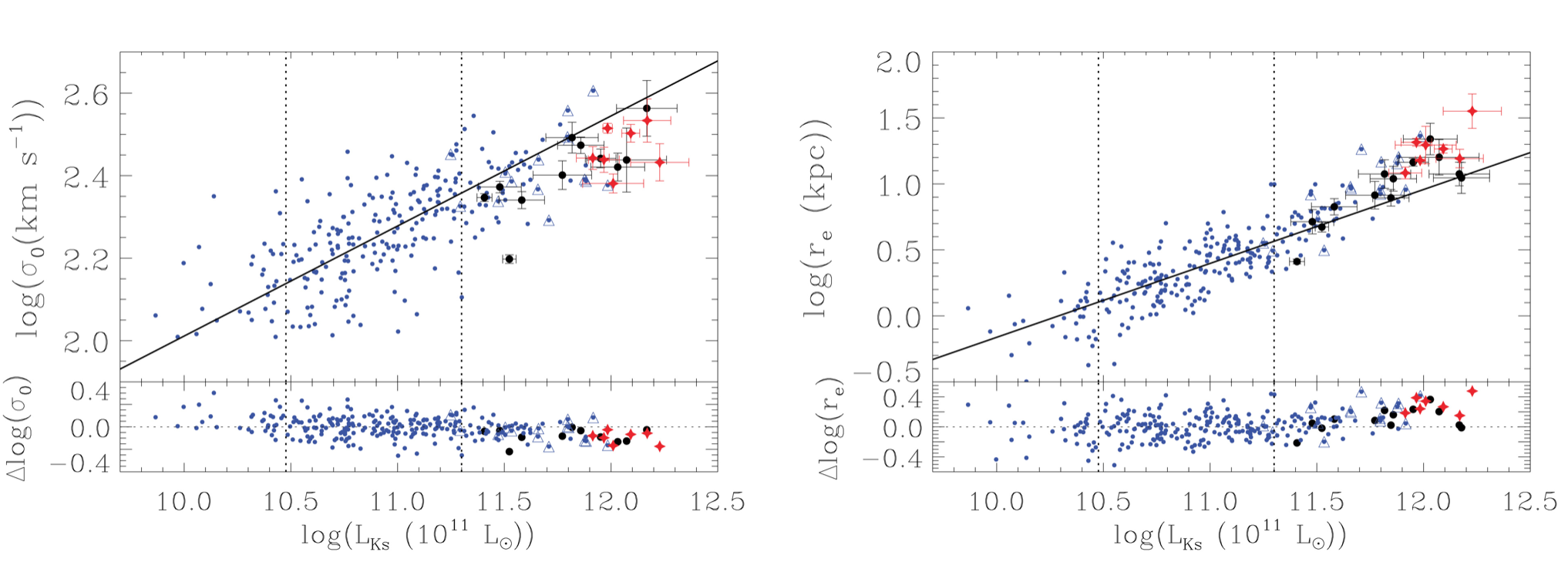}
    \caption{Left panel: distribution of the BCGs of \citet[][ red stars and large black points]{Mendez-Abreu2012} and the early-type galaxies of \citet[][ small blue points]{Pahre1998} in the log $\sigma_0$ vs log $L_{ks}$ plane. The BCGs in the \citet{Pahre1998} sample are marked by blue open triangles. Right panel: same as left panel, but in the log $r_e$ vs. log $L_{Ks}$ plane. The solid line represents the best fit to the galaxies in the luminosity range $3 \times 10^{10} < L_{Ks}/L_\odot < 2 \times 10^{11}$. The bottom panels represent the residuals from the best fit. The original image can be found in \citet{Mendez-Abreu2012} and corresponds to their figures 9 and 10.}
    \label{fig:FP}
\end{figure}

Central galaxies in clusters are a unique class of objects. They are usually the largest and most luminous member galaxies and they lie very close to the peak of the cluster X-ray emission \citep[e.g.][]{Jones1984,Lin2004}. Moreover, in the velocity space they sit near the rest frame velocity of the cluster \citep{Quintana1982,Zabludoff1990,Oegerle2001}. These characteristics imply that the BCGs are located at the minimum in the cluster potential well.
\citet{Zarattini2019} found that there is a dependence of the velocity segregation on the magnitude gap. This result means that BCGs in FGs are located closer to the minimum of the cluster's potential well, when compared to BCGs in non-fossil systems. This difference is not found for satellite galaxies, independent on their mass.

The formation of central galaxies in FGs is thought to be the end result of the group evolution \citep{Ponman1994,Jones2003}. The main actor, in this scenario, would be dynamical friction. However, \citet{Sommer-Larsen2006} suggested that the main difference between FGs and non-FGs has to be found in the initial velocity distribution. In particular, he found that satellite galaxies in FGs should be located on more radial orbits than in non-FGs. This would favour low angular momentum mergers, for which dynamical friction could be more effective \citep{Lacey1993}.

\citet{Mendez-Abreu2012} analysed the photometric properties of central galaxies in FGs. They studied the position of the central galaxies in FGs in the fundamental plane and its projections to explain the formation of these objects (see Fig. \ref{fig:FP}). Central galaxies in FGs results to have large $K$-band luminosities. The $K_s$ luminosity is a good proxy of the total stellar mass since the typical M/L $\sim 1$ for an old stellar population \citep{Bruzual2003}. Thus, it seems clear that central galaxies in FGs are amongst the most massive galaxies known. The left panel of Fig. \ref{fig:FP} shows the correlations between the $K_s-$band luminosities and the central velocity dispersion \citep[Faber-Jakson relation,][]{Faber1976}. Similarly, the right panel of Fig. \ref{fig:FP} shows another projection of the fundamental plane, the $K_s-$band luminosity vs. effective radius ($r_e$). In both relations BCGs in FGs are found slightly outside the correlations: this bend was found for the first time in \citet{Bernardi2011} for early-type galaxies and it can be interpreted in terms of the formation scenario of the BCGs. In particular, if major dissipationless mergers between galaxies are the main mechanisms to build up the mass of the BCGs, the final size is expected to increase, but not its central velocity dispersion; however, if minor dry mergers are the predominant mechanism, they are expected to change both the size and velocity dispersion of the BCGs. Thus, the results presented in \citet{Mendez-Abreu2012} seems to favour the first scenario for these massive galaxies. 
This is also supported by new results obtained from numerical simulations. \citet{Kundert2017} investigated the origin of FGs in the Illustris simulation. With respect to the stellar mass assembly of the BCGs, they found (see their fig. 6) that the accretion is similar at high-redsfhits for fossil and non-fossil systems, whereas a clear difference start to appear at $z \sim 0.3$. From this point, BCGs systematically accreted more mass in FGs than in non-FGs.

However, this formation scenario is in contradiction with the one proposed by \citet{Khosroshahi2006}: in fact, these authors found disky isophotes in their central regions of BCGs, in contrast with most BCGs in non-fossil systems \citep{Rest2001}. This result seems to favour a scenario in which mergers in FGs were rich in gas, thus including large M$^*$ spirals. A similar result was found also by \citet{Eigenthaler2009}, since they found many shells around central galaxies in FGs. These shells are likely formed recently via major mergers of spiral galaxies \citep{Hernquist1992}.

Signs of recent mergers were indeed found by \citet{Alamo2012} while analysing the surface brightness profiles of 3 FGs. However ,this work was mainly focused on the study of globular clusters in FGs. These objects are powerful tools to study galaxy assembly, since they are old and dense enough to survive galactic interactions. \citet{Alamo2012} results seem to point out that globular clusters in BCGs formed in a similar way in fossil and non-fossil systems. In terms of the formation scenario of the BCGs, this can be take as a confirmation that similar processes are at work in the formation of BCGs in fossil and non-fossil systems, although more statistics is needed to generalise this conclusion. A similar result, in terms of the formation scenario, is found also in \citet{Madrid2011} and \citet{Madrid2013}, in which the authors compared the properties of ultra-compact dwarf galaxies (UCDs) in FGs and in the Coma cluster. UCDs are considered the bright-and-massive tail of the globular cluster distribution. Their results showed that UCDs are likely to be a common occurrence in all environments.

The presence of AGNs can also be used to study the formation scenarios of BCGs. In fact, AGNs need major mergers to form, since they use the gas provided by the merger as their fueling mechanism. On the other hand, if no other merger occurred, they are destined to end this fuel and inactivate. \citet{Hess2012} studied the sample of 34 FG candidates from \citet{Santos2007} using radio observations in order to detect the presence of AGNs in FGs. They found that 67\% of these FG candidates contain a radio-loud AGN.   This result seems in contrast with the old formation expected for FGs: in fact, AGN should have run out of fuel since FGs's last major merger. For this reason, \citet{Hess2012} suggested that other mechanisms, such as minor mergers, cooling flows or late time accretion should be invoked to keep the AGNs alive in FGs. However, it is worth noting that the \citet{Santos2007} sample was not pure and about half of the sample was formed by non-fossil systems \citep{Zarattini2014}.


\subsubsection{Central Galaxies: stellar populations properties}
We already mentioned that the formation scenario proposed by \citet{Ponman1994} supposed that FGs formed at high redshift, with few interactions with the large-scale structure along their lives. This would leave enough time for the M$^*$ galaxies to merge with the BCG, thus forming the $\Delta m_{12}$ gap. On the other hand, \citet{Mulchaey1999} suggested the so-called {\it failed group} scenario, in which the BCG is formed as a local over density and no other bright galaxy formed within the group.

However, these two scenarios should also leave clear imprints in the stellar populations of the BCGs. In fact, in the {\it failed group} scenario the central galaxy formed via monolithic collapse, that is expected to create large radial metallicity gradients in the distribution of the stars. On the other hand, in the {\it {\it merging scenario}} the BCG suffered various major mergers, that have the power to erase those gradients, since they mix up the stars and gas during the merging process \citep[e.g.][]{Pipino2010}.

\citet{LaBarbera2009} were the first in using stellar populations to investigate difference between FG's BCGs and regular elliptical galaxies. They used spectra from the SDSS DR4 for their sample of 25 BCGs in FGs and 17 field elliptical galaxies, that act as the control sample. They searched for the single-stellar-population model that best fits the spectra, thus computing mean ages, metallicities, and $\alpha-$enhancement for the two populations of BCGs. The authors showed that no significant difference is found in these paramenters and concludes that BCGs in FGs did not form earlier than the other galaxies.

\citet{Harrison2012} select a sample of 17 FG candidates by combining XMM observations with SDSS DR7 data. They analyse the stellar populations of the BCGs of these systems using SDSS spectra and the Starlight code \citep{CidFernandes2005}. Their results showed no significant differences in stellar star formation rates, age, and metallicities between FGs and their two control samples, one built by optically-selected BCGs, the other with X-ray-selected BCGs.

\citet{Eigenthaler2013} studied the presence of such gradients in age and metallicity for a sample of six BGCs in FGs using longslit spectroscopy from the 4.2m William Herschel Telescope (WHT). They found that the metallicity gradient is flatter with respect to the predictions of the monolithic collapse ($\sim -0.2$ instead of $-0.5$), thus indicating the presence of mergers during the life of the BCGs. On the other hand, the age gradient is, on average, negligible.

\begin{figure*}
    \centering
    \includegraphics[scale=0.5]{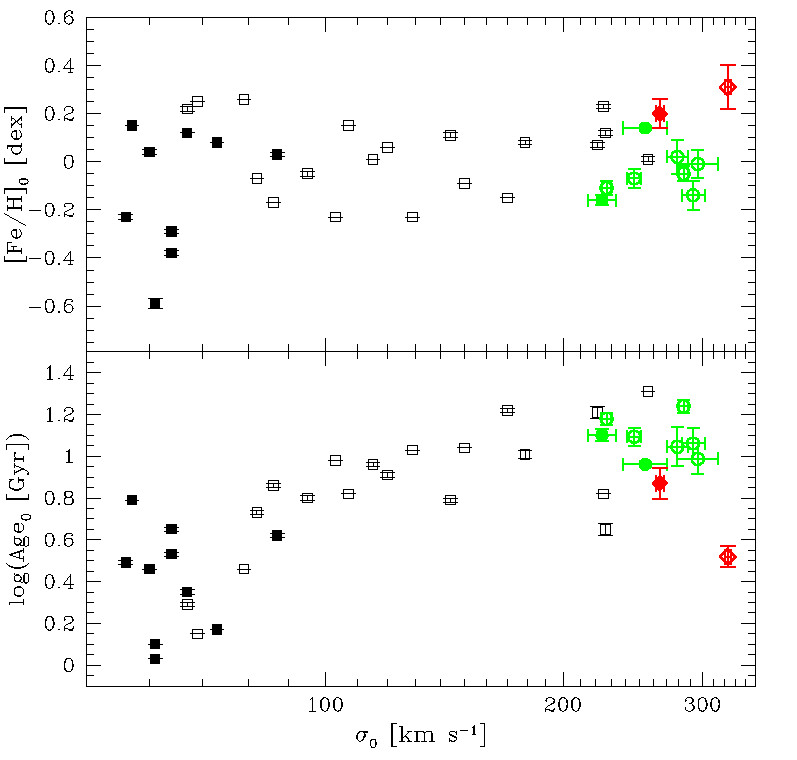}
    \caption{Central metallicity (top panel) and central age (bottom panel) as a function of the central velocity dispersion. Red diamonds are taken from \citet{Corsini2018}, green open circles from \citet{Eigenthaler2013}, green filled circles from \citet{Proctor2014}, open and filled squares are the early-type normal and dwarf galaxies with $\sigma > 50$ km s$^{-1}$ from \citet{Koleva2011}. The original image can be found in \citet{Corsini2018} (see their figure 8).}
    \label{fig:stellar_pop}
\end{figure*}

A similar study was performed by \citet{Proctor2014} for a sample of 2 central galaxies in FGs, using longslit spectroscopy obtained with the 8m Gemini North telescope. The authors found different results for the two BCGs: SDSS J073422.21+265133.9 showed a strong metallicity gradient and a slightly positive age gradient, suggesting a relatively recent episode of stellar formation in the centre. NGC 2484, on the other hand, showed an old stellar population ($\sim 10$ Gyr) and a flat central metallicity, that was interpreted as the evidence of an inside-out stellar formation, at least in the final episode of stellar formation.

\citet{Trevisan2017} studied a large sample of 550 groups to characterise the dependence of the stellar populations of the BCGs and the second brightest galaxies with the magnitude gap. They did not find differences in the distribution of colours, star-formation rates, $\alpha-$enhancement, age, metallicities and star-formation histories in systems with different $\Delta m_{12}$.

\citet{Corsini2018} studied both the stellar populations and radial gradients for a sample of two BCGs in FGs using the 10.4m Gran Telescopio Canarias (GTC) telescope. They confirmed the results of \citet{LaBarbera2009} and \citet{Eigenthaler2013} out to the effective radius. They found an underlying and diffuse older stellar population, with a younger one located near the centre of the galaxies. This was interpreted as the sign of the last major merger with gas, occured $\sim 5$ Gyr ago. \citet{Corsini2018} also found a radial metallicity gradient in agreement with the \citet{Eigenthaler2013} one.

Finally, \citet{Raouf2019} divided a sample of groups from the Galaxy And Mass Assembly (GAMA) survey into relaxed and unrelaxed, using $\Delta m_{12}$ and the luminosity offset as the relaxation indicators. They found that BCGs in unrelaxed systems are bluer, more star forming, and with non-elliptical morphologies than those in relaxed systems. They conclude that the higher rate of recent mergers expected in unrelaxed groups could be responsible for these differences. A similar result was also found in \citet{Pierini2011}. These authors claimed that there are few star-forming galaxies in FGs, making them more mature then coeval and similar mass groups. 

Very recently \citet{Raouf2021} studied the kinematic of gas and stars in 154 central galaxies taken the Sydney-AAO Multi-object Integral field (SAMI) galaxy survey. In particular, they divided this sample into low and high luminosity gap system, with the latter that can be assimilated as FGs. They found that there is a weak statistical difference (at approx. 1$-\sigma$ level) between the magnitude gap and the gas-star kinematics misalignment. In addition, a similar difference was observed between the magnitude gap and the regularity of the stellar rotation of the BCGs. In particular, systems with high magnitude gaps are found to be more regular rotators and with a smaller fraction of gas-star misaligned kinematics.

These studies did not find relevant differences in the stellar populations of central galaxies in fossil and non-fossil systems. The imprint of monolithic collapse is not found, all the observations point towards the creation of FGs via the {\it {\it merging scenario}}, in which the gap is created via major mergers of M$^*$ galaxies. The possibility of the existence of a fossil phase in the life of a cluster is supported also by the presence of younger stellar populations in the centre of galaxy, probably due to recent major mergers with gas. In the top panel of Fig. \ref{fig:stellar_pop} we show the relation between central velocity dispersion and central metallicity for a sample of ten FGs taken from \citet{Eigenthaler2013,Proctor2014,Corsini2018} and compared with the sample of normal and dwarf early-type galaxies of \citet{Koleva2011}. In the lower panel of the same figure, the comparison is done for the central ages of the same samples. It can be seen that in both cases FGs are found in the same correlations as normal early-type galaxies. This plot again confirms that central galaxies in FGs are amongst the most massive known in the Universe.

\subsubsection{Luminosity functions}
\label{sec:LFs}
The luminosity function (hereafter LF) is one of the most powerful tools to study the galaxy population of a group/cluster of galaxies. It is given by the number density of galaxies per luminosity interval and it is usually described parametrically with the Schechter function \citep{Schechter1976}. The main parameters are the characteristic magnitude (M$^*$) and the faint-end slope ($\alpha$). The former describes the bright part of the LF, whereas the latter is related to the dwarf galaxy population. A debate is ongoing on the universality of the LF: in fact, photometric studies found that LFs in clusters are steeper than in the field \citep[$-2.0 < \alpha < -1.8$ in clusters and $-1.5 < \alpha < -1.3$ in the field, see][]{Popesso2006,Blanton2005}. On the other hand, spectroscopic studies found no differences in the $\alpha$ parameters of the field and clusters, finding a general value of $\alpha \sim -1.3$ \citep[][ and references therein]{Aguerri2020}

The study of LFs in FGs was mainly focused on individual FGs, due to their paucity. As a consequence, most of the first results were contradictory. \citet{MendesdeOliveira2006} found, for the FG called RX J1552.2+2013, M$^*=-21.18\pm 0.57$ and $\alpha= -0.77\pm 0.37$ using spectroscopically-confirmed members, or  M$^*=-21.27\pm 0.62$ and $\alpha= -0.64\pm 0.30$ for photometrically-selected galaxies in the $r-$band. In the same year, \citet{Khosroshahi2006} computed $M^*=-20.40\pm0.22$ and $\alpha= -1.23\pm 0.28$ for RX J1416.4+2315, a fossil system with a mass similar to RX J1552.2+2013. Also \citet{Trentham2006} presented an LF for a single FG (NGC 1407), finding $\alpha = -1.35$.
The difference, especially in the faint-end slope, is important. In fact, $\alpha = - 1$ indicates a flat LF, in which the number of dwarf galaxies is not changing with magnitude. On the other hand, in a steeper function like that of \citet{Khosroshahi2006}, the number of dwarf galaxies is rapidely growing and, in a flatter one like that of \citet{MendesdeOliveira2006} it is decreasing. 

\citet{Zibetti2009} also studied the photometric LF of a sample of five FGs. They found a faint-end slope in agreement with the one of regular clusters presented in \citet{Popesso2006}. However, it is worth noting that \citet{Popesso2006} found an upturn at fainter magnitudes, so that their LF can be fitted with a double Schechter function. This behaviour is not found on the already-cited LF of FGs, because none of these are deep enough. However, \citet{Lieder2013} computed a very deep LF for NGC 6482 using spectroscopic data from the Subaru/Suprime-Cam, finding $\alpha= -1.32\pm 0.05$. They did not fit a double Schechter function, however a change in the faint-end slope is present also in their fig. 12.

\citet{Aguerri2011} presented the LF of RX J105453.3+552102, a massive FG at $z=0.5$. They found $M^*=-20.86\pm0.26$ and $\alpha=-0.54\pm0.18$, thus confirming a flatter trend for the dwarf galaxy population. \citet{Adami2012} also computed the LF for two FG, finding that their faint-end slope is relatively flat, but without giving numbers.
Finally, \citet{Aguerri2018} studied the spectroscopic LF of RXJ075243.6+455653, finding $\alpha=-1.08\pm0.33$. 

\begin{figure*}
    \centering
    \includegraphics[scale=0.4]{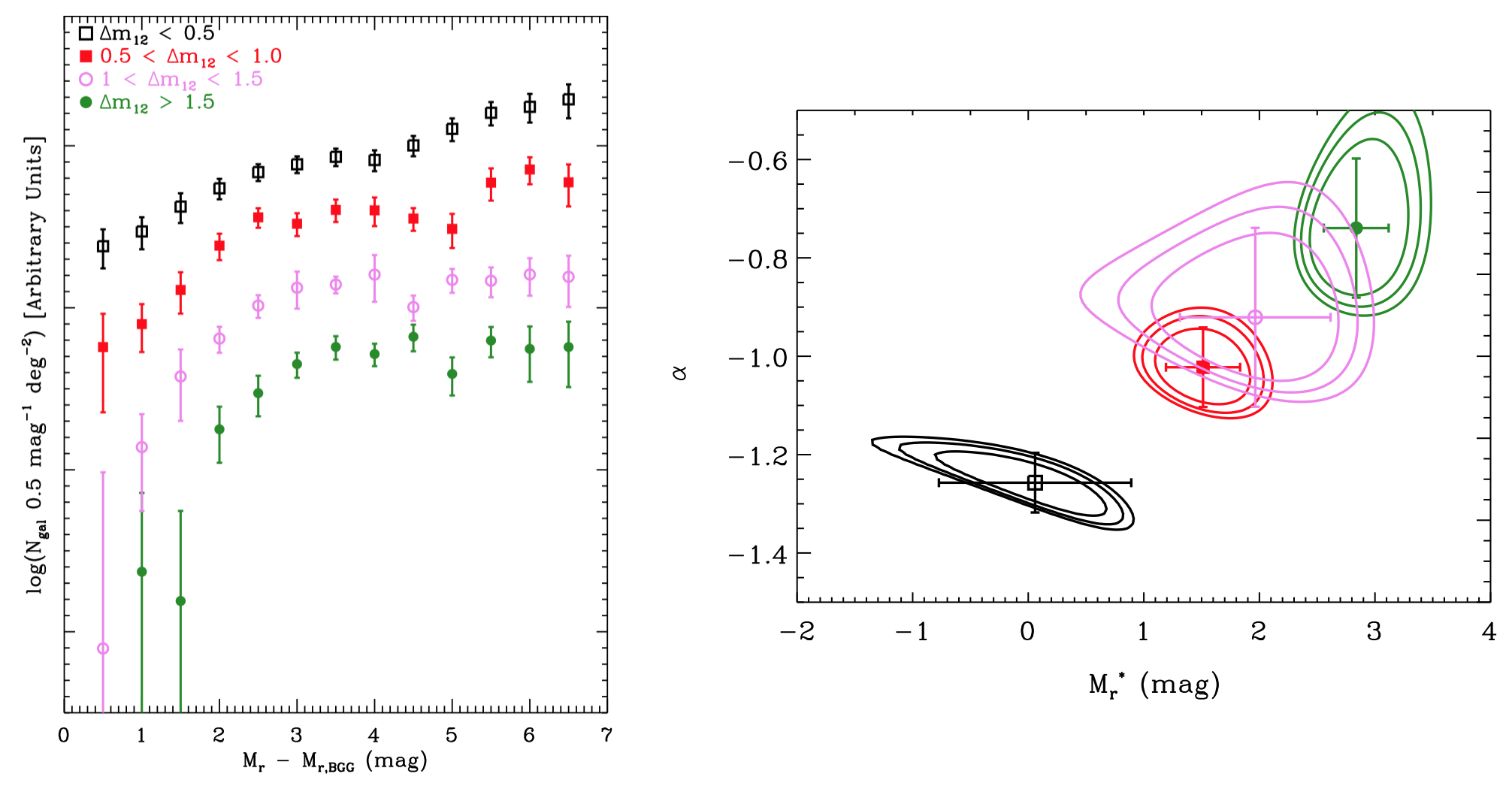}
    \caption{Left panel: {\it relative} LFs for the four subsamples presented in \citet{Zarattini2015}. Empty black squares are systems with $\Delta m_{12} < 0.5$, filled red circles are systems with $0.5 < \Delta m_{12} < 1.0$, empty violet circles are systems with $1.0 < \Delta m_{12} < 1.5$, and filled green circles are systems with $\Delta m_{12}$ > 1.5. Right panel: Uncertainty contours for the Schechter fits of LFs in the left panel. Contours represent 68\%, 95\%, and 99\% c.l. and the colour and symbol codes are the same as in left panel. Original image can be found in \citet{Zarattini2015}, in particular left panel correspond to their figure 7, right panel to figure 8.}
    \label{fig:LFs}
\end{figure*}

The first systematic study of the dependence of the LF on the magnitude gap was presented in \citet{Zarattini2015}. The authors selected $\sim 100$ clusters and groups spanning a wide $\Delta m_{12}$ range and dividing their analysis in four bins of $\Delta m_{12}$. Their study was based on a hybrid method for computing the LF, in which the bright part was treated as a quasi-spectroscopic LF, whereas in the faint end photometric data were dominant. The authors computed a classical LF and one in which the magnitudes of each systems are referred to the magnitude of the central galaxy (e.g. $M_r - M_r,BCG$, called {\it relative} LFs). The latter permits to compare directly the differences due to the magnitude gap and the authors found that this technique offers the best results for highlighting the differences between their four subsamples. These {\it relative} LFs are shown in the left panel of Fig. \ref{fig:LFs}. \citet{Zarattini2015} found that both M$^*$ and $\alpha$ changes with the magnitude gap. In particular, systems with $\Delta m_{12} < 0.5$ have the brightest M$^*$ and the steepest $\alpha$ slope, whereas systems with $\Delta m_{12} > 1.5$ have the faintest M$^*$ and the flattest $\alpha$. The differences are larger than 3$\sigma$ between the two most-extreme cases and the trend with $\Delta m_{12}$ is clearly visible in the right panel, where the value of the {\it relative} M$^*$ and $\alpha$ are shown for the four subsamples. 

The discussion is usually focused on the faint end of the LF, since differences in the bright part between fossils and non-fossils can be easily explained with the same mechanisms that are responsible for the creation of the magnitude gap. Moreover, the presence of the gap itself as a selection criteria implies differences in the bright part of the LFs.

On the other hand, the debate on the differences in the faint-end slope is more complex. In fact, dwarf galaxies can not be merged into the BCG in a reasonable time, since the merging time scale is inversely proportional to the mass of the satellite. A possible explanation is that dwarf galaxies are located in radial orbits passing close to the centre of the cluster. This will result in their disruption, accounting for a part of the missing population. This explanation was also invoked as a possible reason for the high merging rate of FGs. It can also justify the formation of the magnitude gap, since \citet{Lacey1993} showed that the merging timescale for satellite on radial orbits is shorter than for tangential ones.
Another possibility is that FGs could lack dwarf galaxies for differences in their accretion time. In fact, \citet{Aguerri2018} found that the large-scale environment of FGS03 is very rich, so the flat LF found could be explained if the dwarf population are still trapped in nearby groups, awaiting to be merged with the FG. In this context, FGs would be systems in early stages of their mass assembly. We will discuss in more details the large-scale structure of FGs in Sect. \ref{sec:large_scale}.

Differences in the accretion history of dwarf galaxies were indeed found also in \citet{Kanagusuku2016} while studying the bright and dwarf galaxy populations in fossil and non-fossil clusters in the Millennium simulation. They found that FGs had a denser dwarf population at early time ($z > 0.7$), then the trend reverses ($0.5 < z < 0.3$), and finally it become similar at $z=0$.

To finally solve the issue of the observed lack of dwarf galaxies in FGs, deep and extended spectroscopy would be needed. We will discuss in Sect. \ref{sec:future} the expected impact on this topic of the next-generation spectroscopic surveys.

\subsubsection{Galaxy substructures}

If the {\it old and relaxed} model is correct, one should expect to find a smaller amount of galaxy substructures in FGs than in non-FGs. The only study on this topic was done in \citet{Zarattini2016}. They analysed the sample of 34 FG candidates of the FOGO project to compute the fraction of FGs with signs of substructures using a variety of methods. In particular, the entire sample was studied with a two-dimensional approach, able to detect substructures in the projected space of the cluster. Moreover, for a subsample of candidates for which an extended spectroscopic follow up was available, they also applied a series of one- and three-dimensional tests (e.g. the Dressler-Schectman test). 

\citet{Zarattini2016} results depend critically on the adopted tests, but the comparison with a control sample shows that galaxy susbstructures are present in a similiar fraction in fossil and non-fossil systems. 
This presence of substructures in FGs is hardly compatible with an old formation, followed by a passive evolution, with no major interactions with the surrounding large-scale structure. Indeed, the small number of genuine FGs in the sample prevent to reach a definitive conclusion.

\subsubsection{Large scale environment}
\label{sec:large_scale}

Differences in the large-scale environment and in the way in which it interacts with FGs were invoked as a possible cause of the different evoulution of FGs and non-FGs \citep{D'Onghia2005,Ponman1994,Diaz-Gimenez2011}. Observational results are scarce on this topic, since only few individual FGs were studied so far.

\citet{Adami2007} studied the large-scale structure around the FG RXJ1119.7+2126 using spectroscoic data. They conclude that this FG is located at the centre of a low galaxy density bubble.

\citet{Pierini2011} found controversial results in their analysis of two FGs. One of the two is found in an isolated environment, whereas the second one is located in a dense environment, with 27 other groups or clusters in the surroundings.

Also \citet{Adami2012} studied the environments of other three FGs using photometric and spectroscopic data. They found that one system (1RXS J235814.4+150524) is in a poor environment, though its galaxy density map shows a clear signature of the surrounding cosmic web. The second FG (RX J1119.7+2126) is very isolated, whereas the third one (NGC 6034) is embedded is a very rich environment.

Finally, \citet{Diaz-Gimenez2011} analysed the large-scale structure in FGs and non-FGs from both the theoretical and observational points of view. We already mentiond their theoretical results in Sect. \ref{sec:theory}, here we focus on their observational tests. In fact, they used four FGs selected from \citet{Voevodkin2010} and coming from the 400d cluster survey \citep[][]{Burenin2007} and a control sample of non-FGs from the same survey. Their observational results confirmed the peak found in numerical simulations in the local density profile of galaxies around groups as a function of the normalised group-centric distance. This peak, located at about $2.5 r/r_{vir}$ at $z = 0$, is more prominent in FGs. However, the difference is clearer in numerical simulations than in observations, probably again due to the small sample of FGs available.

More extended studies on larger samples are thus generally needed to confirm if FGs are characterised by a special large-scale environment.

\section{Past and future of fossil systems evolution}
\label{sec:other}

If FGs are the end product of groups/clusters evolution, a question should naturally arise: are their progenitors regular groups/clusters or do they belong to some particular class? A possible answer, that we will discuss in the first part of this section, is that the progenitors could be found in compact groups. In fact, in these systems various bright galaxies are cooped up in a small area, making them ideal candidates for fast and efficient mergers.

However, \citet{vonBenda2008} suggested that the fossil status may be only a transitional phase in the life of a regular cluster. If this is the case, there is no need to find a special category of progenitors, since the acquisition of the fossil status could happen to any group/cluster in the period between the last major merger and the subsequent arrival of another bright galaxy from the cosmic web. We will discuss this topic from an observational point of view in the second part of this section.

\subsection{Compact and loose groups as progenitors of fossil systems}
The discussion on the progenitors of FGs started even before the actual discovery of these systems. In fact, \citet{Ponman1993} suggested that compact groups were the result of orbital decay in larger systems and that they should culminate in a final merger, in which all the bright galaxies would merge at the centre of the system. They also added that a new category of ``fossil groups'' were awaiting discovery in the ROSAT all-sky survey. A year after this claim, they announced the discovery of the first FG \citep{Ponman1994}.

Since then, many studies looked for FG's progenitors, using a variety of techniques. \citet{Miles2004} studied a sample of 25 clusters in the X-ray, finding that some of them are dimmed in luminosity. Their interpretation of the result was that, according to a specific toy simulation, groups with dimmed X-ray luminosity have lower velocity dispersion. This would lead to the formation of FGs, since low-velocity encounters between massive galaxies are the most efficient in terms of merging time scale. This result was confirmed also by the analysis of a compact group at $z=0.22$ studied in \citet{MendesDeOliveira2007}. In fact, these authors showed that this system has many characteristic in common with FGs and, in particular, the merging of the four brightest members would lead to the formation of a BCG of $M_r \sim -23$, a typical value of central galaxies in FGs. A similar result was also found in the study of galaxy pairs \citep{Grutzbauch2009}. In this case, the author claims that E+S pairs could be the last step in the formation of FGs, thus a sort of transition between compact groups and FGs. Also \citet{Pierini2011} found evidences that compact groups are favoured as progenitors over the early assembly. On the other hand, \citet{Yoshioka2004} found a M/L ratio for FGs that is too high if compared with compact groups, thus claiming that these systems are not the ideal progenitors.

A novel technique to search for FG's progenitors is the use of strong gravitational lensing in galaxy groups. \citet{Johnson2018a} investigated the fraction of FGs in lensed and non-lensed galaxy groups, finding that the fraction of FGs is larger in lensed systems (13\% versus 3\%). They also identified 12 possible FG progenitors, that were later investigated in details in \citet{Johnson2018b} using Chandra and the Hubble Space Telescope. Their results showed that the X-ray temperatures of the candidate progenitors are higher than those of the control sample. They also find hints of differences in the LFs of FGs and non-FGs, but these differences are erased when BCGs are removed. Finally, \citet{Schirmer2010} studied the strong-lensed FG J0454-0309, that is found behind a well-studied non-fossil poor cluster. Their analysis support a scenario in which the fossil system is falling into the poor cluster and where the central galaxy of the FG will become the brightest galaxy of the new system.

Another approach was proposed by \citet{Tovmassian2010}: they compared the $K-$band absolute magnitudes of BCGs in regular clusters and FGs, finding that the latter are systematically fainter. The author concluded that FG progenitors are likely poor groups. Moreover, it is interesting to note that \citet{Tovmassian2006} studied ``{\it the properties of Hickson's compact groups and of the Loose Groups within which they are Embedded}''. In this work, no link with FGs is suggested, but the result could be seen under a different light after the \citet{Tovmassian2010} study, confirming the idea that poor compact groups could be the ideal progenitors of FGs.  However, this result apparently collides with what was presented in Sect. \ref{sec:galaxies}, where we cited various studies claiming that BCGs in FGs are amongst the most-massive galaxies in the Universe. On the other hand, \citet{Farhang2017} analysed the mass assembly histories fo compact and fossil systems in the Millennium simulation and associated semi-analytical models. They found that only 30\% of FGs could originated from compact groups. They conclude that most of the fossil and compact groups follow different evolutionary paths.

Finally, \citet{Buote2018} suggested that compact elliptical galaxies (CEGs) surrounded by an X-ray halo should be considered as FGs. In fact, detailed X-ray observations of two of these systems found that mass and entropy profiles and concentration are compatible with other FGs studied in the literature. In this case, the progenitors are expected to be the so-called ``red-nuggets'', compact galaxies found a $z \sim 2$ \cite{Ferre-Mateu2017}. This case is indeed peculiar: the authors did not study FGs in order to find their progenitors, but better suggested that a new type of galaxy should be considered in the fossil category.

\subsection{Transitional fossil phase}
We already mentioned in Sect. \ref{sec:theory} that \citet{vonBenda2008} suggested that FGs are only a transitional phase in the life of a regular group/cluster. They claimed that this phase would happen just after a major merger and before other bright galaxies are accreted to the group. An example of such a process is found in \citet{Irwin2015}: the Cheshire Cat galaxy group is formed by two smaller groups, dominated by one bright galaxy each. These groups are experiencing a line-of-sight merger that will end up in approximately one Gyr with the merging of the two structures. The authors suggested that the resulting structure will be a massive fossil group, dominated by a large $M_r = -24$ galaxy.

A similar case is the one presented in \citet{Aguerri2018}: RX J075243.6+455653 was found to actually accomplish the fossil definition of $\Delta m_{12} > 2$ within half the virial radius, however another galaxy almost as bright as the BCG is found just outside that radius. Depending on its orbit, RX J075243.6+455653 became fossil in the very last part of its life or, in the opposite case, it will become non-fossil in the near future.

The existence of a fossil phase may thus explain some of the controversial results presented along this review. It is possible that the observational definition based on the magnitude gap alone is not sufficient to clearly separate the population of real FGs to that of non-FGs dominated by a massive central galaxy. This would confirm the results of \citet{Raouf2014}, since the authors suggested that other observational quantities (like the luminosity of the BCG and its separation from the luminosity centroid of the group) should be used to create a sample dominated by purely old FGs. We thus suggest to start using this new definition in the search for FGs as a way for creating a sample of old systems. However, using these additional observational constraints could dramatically reduce the number of identified systems. In Sect. \ref{sec:newsample} we will give a list of the most-secure FGs up to date: only one out of 18 FGs for which the absolute magnitude of the BCG is available will survive the application of the \citet{Raouf2014} criterium ($\Delta m_{12} > 2.0$ and $M_{r,BCG} > -22.5$).

\section{Discussion and conclusions}
\label{sec:conclusions}
In this section we propose a sample of genuine FGs that can be the starting point for new follow ups of these objects. Then we discuss what we presented along the review and draw our general conclusions.

\subsection{Sample of genuine fossil groups}
\label{sec:newsample}

In Table \ref{tab:sample} we present a list of confirmed FGs in the literature. The goal is to offer to the reader a sample as pure as possible for future follow ups. The list is probably not complete, but we did our best to select FGs applying a rigorous criterium on the $\Delta m_{12}$ parameter. In particular, we consider as fossils those systems with $\Delta m_{12} \ge 2.0$ within half the (projected) virial radius. Moreover, we exclude FGs for which membership was done using a fixed cut in $\Delta z$, except when no ambiguous galaxy was found within half the (projected) virial radius. The magnitude gap between the first and fourth brightest galaxies ($\Delta m_{14}$) is also given, when available, but it was not used for the selection, since it was computed only in the most recent studies. 
The absolute $r-$band magnitude of the BCG is included, when available, to simplify the application of the \citet{Raouf2014} criterium. Moreover, we also list the mass of the system, when available. However, we note that the mass is computed in a very inhomogeneous way (different methods and radii), our goal is to offer an at-a-glance reference to the reader.
Finally, the redshift is given for all FGs listed in the table.

We note that, if contradictory information are available, we always choose to apply the \citet{Jones2003} criteria in the most severe way. For example, in \citet{Proctor2011} the sample of \citet{Miller2012} was studied in more details, computing $r_{200}$ in two different ways: one obtained from weak lensing analysis and the other from X-ray data. The latter is found to be $\sim 50\%$ larger than the former. As a consequence, the number of FGs found using the smallest radius is 10, a number that reduces to 3 if the largest radius is used. In order to provide the cleanest sample of genuine FGs, we include in Table \ref{tab:sample} only the 3 obtained with the largest $r_{200}$, citing only \citet{Proctor2011}, even if most of the candidates were also present in \citet{Miller2012}. This choice is done in order to direct the reader to the most up-to-date and/or relevant information.

\begin{table*}[!htbp]
\begin{small}
\caption{A non-exhaustive list of confirmed FGs for which at least $\Delta m_{12} \ge 2$ (and eventually $\Delta m_{14} \ge 2.5$) is computed within half the virial radius in the literature.} 
\label{tab:sample}
\centering
\begin{tabular}{l|c|c|c|c|c|c}
\hline

\multicolumn{1}{c}{Name} &  \multicolumn{1}{c}{$\Delta m_{12}$} & \multicolumn{1}{c}{$\Delta m_{14}$} & \multicolumn{1}{c}{$M_{r,BCG}$} & 
\multicolumn{1}{c}{z} &   
\multicolumn{1}{c}{Mass} & \multicolumn{1}{c}{Reference} \\  
\multicolumn{1}{c}{} &  \multicolumn{1}{c}{} &  \multicolumn{1}{c}{} & \multicolumn{1}{c}{} & \multicolumn{1}{c}{} &   
\multicolumn{1}{c}{[$10^{14}$ M$_\odot$]} & \multicolumn{1}{c}{} \\  

\hline
RX J1340.6+4018*  & 2.3 & / & / & 0.171 & 0.28 &  \citet{Ponman1994} \\
RXJ1119.7+2126 & > 2.5 & / & $-22.8$ & 0.061 & / & \citet{Jones2003} \\
RXJ1331.5+1108 & 2.0 & / & $-23.6$ & 0.081 & / & \citet{Jones2003} \\
RXJ1416.4+2315 & 2.4 & / & $-25.0$ & 0.137 & / & \citet{Jones2003} \\
RXJ1552.2+2013** & 2.3 & / & $-24.7$ & 0.135 & / & \citet{Jones2003} \\
NGC 6482 & 2.06 & / & $-22.7$ & 0.0131 & $ 0.042 $ & \citet{Khosroshahi2004} \\
ESO 3060170 & 2.61 & / & $-24.4$ & 0.0358 & $1-2$ & \citet{Sun2004} \\ 
UGC 842 & 2.99 & / & $-23.0$ & 0.045 & $ 0.4 $ & \citet{Voevodkin2008} \\
AWM 4 & 2.23 & / & / & 0.0317 & $1.4 $ & \citet{Zibetti2009} \\
J0454-0309 & 2.5 & / & $-24.1^\dagger$ & 0.26 & $0.75-0.90 $ & \citet{Schirmer2010} \\
RXC J0216.7-4749 & > 2.21 & / & / & 0.064 & $0.8 $ & \citet{Democles2010} \\
CXGG 095951+0140.8 & 2.10 & / & $-24.9^\dagger$ & 0.372 & $0.95$ & \citet{Pierini2011} \\
CXGG 095951+0212.6 & 2.32 & / & $-23.9^\dagger$ & 0.425 & $0.19$ & \citet{Pierini2011} \\ 
SDSS J0906+0301 & 3.09 & / & / &0.1359 & $1.3 $& \citet{Proctor2011}\\
SDSS J1045+0420 & 2.00 & / & / & 0.1539 & $2.2 $& \citet{Proctor2011}\\
1RXS J235814.4+150524 & > 2.0 & / & / & 0.178 & / & \citet{Adami2012} \\
DMM2008 IV & 2.4 & 3.0 & / & 0.0796 & / & \citet{Harrison2012} \\
WHL J083454.9+553421 & 2.4 & 3.0 & / & 0.2412 & / & \citet{Harrison2012}\\
A0963 & 2.2 & 2.7 & / & 0.2056 &  / & \citet{Harrison2012}\\
A1068 & 2.3 & 3.1 & / & 0.1381 & / & \citet{Harrison2012}\\
BLOX J1230.6+1113.3 ID & 2.1 & 3.5  & / & 0.1169 & / & \citet{Harrison2012}\\
XMMXCS J123338.5+374114.9 & 2.6 & 3.2 & / & 0.1023 & / & \citet{Harrison2012} \\
ZwCl 1305.4+2941 & 2.6 & 3.1 & / & 0.2406 & / & \citet{Harrison2012} \\
MaxBCG J197.94248+22.02702 & 2.1 & 2.7 & / & 0.1715 & / & \citet{Harrison2012} \\
XMMXCS J134825.6+580015.8 & 2.0 & 2.6 & / & 0.1274 & / & \citet{Harrison2012}\\
XMMXCS J141657.5+231239.2 & 2.8 & 3.1 & / & 0.1159 & / & \citet{Harrison2012}\\
XMMXCS J160129.8+083856.3 & 2.4 & 3.1  & / & 0.1875 & / & \citet{Harrison2012}\\
FG12 & > 2.0 & / & / & 0.089 & 0.6 & \citet{LaBarbera2012}\\
FGS02 & > 2.21 & > 2.28 & $-25.0$ & 0.23 & 18.7  &\citet{Zarattini2014} \\
FGS03 & 2.09 & 2.55 & $-22.6$ & 0.052 & 0.42  &\citet{Zarattini2014} \\
FGS08 & > 2.12 & > 2.17 & $-24.2$ & 0.409 & / &\citet{Zarattini2014} \\
FGS10 &  2.12 & 2.24 & $-25.3$ & 0.468 & 8.32  &\citet{Zarattini2014} \\
FGS20 & 2.17 & > 2.46 & $-23.6$ & 0.094 & 1.63  &\citet{Zarattini2014} \\
FGS28 & > 3.28 & > 3.68 & $-21.3$ & 0.032 & / &\citet{Zarattini2014} \\
2PIGG 2515 & 3.4 & / & $-23.4$ & 0.062 & 0.37  & \citet{Khosroshahi2014}\\
2PIGG 2868 & 2.5 & / & $-23.1$ & 0.067 & 0.25  & \citet{Khosroshahi2014}\\
\hline
\end{tabular}
\end{small}
\begin{minipage}{\textwidth}
{\small Notes. Column (1): System name, as presented in the cited publication. Column (2): Magnitude gap between the two brightest member galaxies. Column (3): Magnitude gap between the first and fourth brightest member galaxies. Column (4): $r-$band absolute magnitude of the BCG. (5): redshift. (6): Mass. Column (7) Reference paper. 

It is worth noting that we only cite minimal references for each FG and the same object can be also found in other publications. Moreover, we only cite systems for which $\Delta m_{12}$ is strictly larger than 2, in order to propose a sample of genuine fossil systems.

* This system is disqualified as fossil in \citet{MendesdeOliveira2009} using $i-$ and $g-$bands.

** This system is disqualified as fossil in \citet{Zibetti2009}. 

$^\dagger$ Computed in the $i-$band.}
\end{minipage}
\end{table*}

It is worth noting that some famous FGs are excluded from the list. As an example, we discuss the prototype of this category, NGC 1132 \citep{Mulchaey1999}, for which we were not able to find $\Delta m_{12}$. There are indeed information on the fainter galaxies \citep[e.g.][]{Dong-woo2018}, but not a clear computation of the magnitude gap. However, \citet{Dong-woo2018}, claimed that the second brightest galaxy is NGC 1126, a spiral galaxy located at $8.4$ arcminutes or 230 kpc in projection. The virial radius of this group is estimated to be $r_{200} \sim 800$ kpc, thus this galaxy should be inside 0.5 $r_{200}$. The difference in the velocity space is $\Delta v = 438$ km s$^{-1}$, as obtained using the Nasa Extragalactic Database (NED), so the two galaxies can be part of the same group, although a precise dynamical study should be done to confirm the membership.
In \citet{Dong-woo2018} NGC 1126 is described as ``seven times faniter in B''. We check in the SDSS DR16 the magnitudes of both NGC 1132 and NGC 1126: the former has $m_r = 12.20$, the latter $m_r = 14.01$, leading to a $\Delta m_{12} = 1.81$, rejecting it as a genuine FG. However, as we already mentioned, an accurate study of the membership of NGC 1126 to the group of NGC 1132 should be done and errors in SDSS magnitudes can not be excluded (NGC 1126 is flagged with ``unreliable photometry'', as many other bright galaxies, mainly due to an over estimation of the sky around bright objects).
Concluding, with this example we aim at demonstrating that also the definition of the most commonly-accepted FGs may be not rigorous, or may need deeper studies to include them in a sample of genuine FGs. 

\subsection{Conclusions and future prospects}
\label{sec:future}
Along this review we analysed the most-studied topics on FGs. The aim of these studies was to test the so-called {\it merging scenario}, which predicts that fossil systems formed earlier than non-fossils, having enough time to merge all the bright galaxies with the BCG and remaining somewhat isolated from the cosmic web (e.g. they did not receive other bright galaxies from the merging with other groups/clusters). However, the general framework that can be obtained from this review is that FGs probably formed and evolved in a similar way as non-FGs. In particular, we show that early differences reported in global properties such as the scaling relations and M/L ratios of the halos of fossil systems can be reconciled when homogeneous datasets are used.  Probably, an analogous result would be obtained for the differences observed in the mass and the entropy profiles in some individual systems. A more homogeneus and large sample is required to be analysed in this case. Moreover, no differences are found in the fraction of galaxy substructures identified in FGs and non-FGs. This again indicates that the halos of FGs are not significantly older than those from non-FGs.

The central galaxies in fossil groups show similar stellar ages and metallicities than BCGs in the center of non-fossil systems. In addition, the location of these galaxies in the fundamental plane and its projections indicate a formation process driven by dissipanionless mergers in a similar way as other bright early type galaxies.

The similarities found in the formation of fossil and non-fossil systems seem to indicate that the large magnitude gap could just be a transient phase in the evolution of groups and clusters, as reported by different numerical simulations. This magnitude gap would be more connected with recent major mergers rather than with an old formation.

If this is the case, one should find an explanation for those differences that can not be reconciled with inhomogeneities in the data.  We already mentioned that \citet{Sommer-Larsen2006} proposed that more radial orbits for galaxies in fossils could be responsible for the formation of the gap. This idea is supported also by \citet{Lacey1993}: the merger timescale with the central halo is shorter for M$^*$ galaxies on radial orbits than for galaxies on tangential orbits (see their eq. 4.2). From an observational point of view, hints of this difference are found in \citet{Zarattini2021}. The authors studied the orbital structure of a sample of $\sim 100$ groups and clusters, dividing them in four bins of $\Delta m_{12}$. Their larger magnitude gap bin ($\Delta m_{12} > 1.5$) shows the presence of radial orbits in the external regions ($0.8 - 1 \,r_{200}$), that is not found in the other three bins, all with $\Delta m_{12} < 1.5$. However, the results should be confirmed with a larger sample of genuine FGs, as we already mentioned along this review for a significant part of the discussed topics.

The other main topic that remains open is the difference found in the faint-end slope of FGs LFs. This is difficult to explain within the current models of formation and evolution of clusters. In fact, most of the studies points toward a sort of {\it global} value for the faint-end slope in clusters and groups. It is worth noting, however, that the majority of the studies of LFs in FGs used photometric data and that, in the literature, significant differences were found even in regular clusters when only photometric data were used. Thus, the next step in this discussion awaits the use of large spectroscopic datasets, that will become available with the next generation of astronomical instruments (e.g. WEAVE, 4MOST, DESI). 
However, we can tentative say that the presence of radial orbits could give an answer also to the problem of the faint-end slope of the LF: in fact, if massive galaxies on radial orbits have a shorter timescale to be merged within the central galaxy, dwarf galaxies could be more easily disrupted on such orbits, if they pass near the BCG \citep{Smith2015}. 
Another possible explanation for the differences in the galaxy populations between FGs and non-FGs could be found in the surrounding environment, since hints of different large-scale structures are found in some individual studies. In particular, we can not exclude that FGs are still in the process of accreting dwarf galaxies, but the general picture remains to be clarified.

The creation of a large and strict sample of genuine FGs and a homogeneous follow up will be the key of the characterisation of FGs in the near future. For this reason we gave in Sect. \ref{sec:newsample} a table with the most-secure FGs to date. The computation of the $\Delta m_{12}$ is not a real issue with surveys like SDSS, DES, or Pan-STARRS1, already available for 3/4 of the sky.  The arrival of new facilities will be useful for the confirmation of the FGs candidates found with these photometric surveys.  New X-ray data will be available with the new all-sky surveys like eROSITA (we refer the reader to the companion review of \citet{Eckert2021} for a detailed description on the impact of eROSITA and other X-ray surveys on the study of galaxy groups). In addition,  the spectroscopic follow up will be possible with extended spectroscopic surveys like WEAVE, 4MOST, and DESI or with precise photometric redshifts surveys like J-PAS. The firm identification of at least 50/100 FGs will be the main scientific goals in this field for the next decade. This will be easily achieved in the near future, since eROSITA is exepcted to find $\sim 10^5$ groups/clusters. For comparison, the REFLEX cluster catalogue has $\sim 1500$ groups/clusters. Most of these new clusters will have dedicated spectroscopic follow ups with the next generation multi-object spectrographs.

\acknowledgments{The authors thanks MNRAS, A\&A, and the AAS, together with the authors of the corresponding publications, for granting permission for using images published in their journals. SZ is funded by Padua University grant ARPE-DFA-2020. JALA was founded by the project AYA2017-83204-P.}

\reftitle{References}

\externalbibliography{yes}
\bibliographystyle{Definitions/aa}
\bibliography{bibliografia}

\end{document}